\documentclass[12pt]{iopart}

\pdfoutput=1
\expandafter\let\csname equation*\endcsname\relax
\expandafter\let\csname endequation*\endcsname\relax
\usepackage[ansinew]{inputenc}
\usepackage{color}
\usepackage{gensymb}
\usepackage{graphicx}
\usepackage{amsmath,amsfonts,amssymb}
\usepackage[bottom,hang]{footmisc}
\usepackage{graphicx}
\usepackage{units}
\usepackage{textcomp}
\usepackage[numbers,sort&compress]{natbib}

\newcommand{\C}{$\text{C}_{\text{60}}$}
\newcommand{\hex}{$h$-$\text{C}_{60}$}
\newcommand{\pen}{$p$-$\text{C}_{60}$}

\newcommand \eg {\emph{e.\,g.}}

\begin{document}

\title{Interactions between two \C\ molecules measured by scanning probe microscopies}

\author{Nadine Hauptmann$^{1,4}$, C\'{e}sar Gonz\'{a}lez$^2$, Fabian Mohn$^3$, Leo Gross$^3$, Gerhard Meyer$^3$, Richard Berndt$^1$}
\address{$^1$Institut f\"{u}r Experimentelle und Angewandte Physik, Christian-Albrechts-Universit\"{a}t zu Kiel, D-24098 Kiel, Germany\\}
\address{$^2$Departamento de Fisica, Facultad de Ciencias, 33006. Universidad Oviedo, Spain\\}
\address{$^3$IBM Research -- Zurich, CH-8803 R\"{u}schlikon, Switzerland\\}
\address{$^4$Institute for Molecules and Materials, Radboud University, 6500 GL Nijmegen, The Netherlands}

\ead{n.hauptmann@science.ru.nl}

\begin{abstract}
\C-functionalized tips are used to probe \C\ molecules on Cu(111) with scanning tunneling and atomic force microscopy.
Distinct and complex intramolecular contrasts are found.
Maximal attractive forces are observed when for both molecules a [6,6] bond faces a hexagon of the other molecule.
Density functional theory calculations including parameterized van der Waals interactions corroborate the observations.
\end{abstract}

\maketitle

\section{Introduction}
The interactions between two molecules in nanometer proximity via forces and via charge transfer are important in a wide range of research areas.
Scanning probe microscopy provides an opportunity to investigate these interactions in some detail.
Molecules may be immobilized on a substrate and on the tip of the microscope and their relative positions may be controlled with picometer precision.
The orientation of the molecule on the surface may be determined from conventional imaging.
When a suitable ''micro tip'' is prepared on the substrate, imaging of the molecule at the tip can be achieved as well~\cite{Schull2009,Chiutu2012}.\\
By combining atomic force and scanning tunneling microscopy (AFM/STM), both force and charge transport may be probed.
Combined AFM and STM employing metallic tips has been used to study various molecules, \emph{e.~g.}, diatomic~\cite{Sun2011} and planar~\cite{Gross2009,Gross2010,Fournier2011, Mohn2011, Pavlicek2012,Hamalainenn2014,Oteyza2013,Albrecht2013, Yang2014} molecules, functional molecular complexes~\cite{Hauptmann2015}, as well as \C~\cite{Gross2012,Pawlak2011, Hauptmann2012,Chiutu2012}.
Using a \C\ molecule at the tip and \C\ molecules on the surface, we previously investigated how the charge transport \cite{Schull2009} and the force \cite{Hauptmann2012} vary as the tip is brought closer to the sample.
As a result, the short-range force around the point of maximal attraction could be determined as a function of the intermolecular distance.
Owing to their rigidity \C\ tips may also be used for lateral scanning~\cite{Nishino2005,Schull2009,Chiutu2011,Hauptmann2012}.
This enables measurements in the distance range near mechanical contact that extend the previous one-dimensional measurements to the lateral dimensions.\\
Here we use \C-functionalized tips to laterally probe \C\ molecules by combined STM/AFM measurements together with density functional theory (DFT) calculations taking van der Waals (vdW) interactions into account.
Maximal attraction is found when a [6,6] bond of one molecule faces a hexagon of the other molecule and vice versa.

\section{Experimental and theoretical methods}
The experiments were performed with a homebuilt combined STM/AFM in ultrahigh vacuum at a temperature of $5\,\text{K}$.
The atomically flat Cu(111) surface was used as substrate, which was prepared by repeated sputtering and annealing cycles.
Submonolayer amounts of \C\ were then deposited onto the sample by sublimation at room temperature.
Subsequent annealing to $\approx 500\,\text{K}$ led to a well-ordered $4\times 4$ structure of \C~\cite{Hashizume1993,Pai2004,Wang2004,Pai2010}.
Isolated \C\ molecules were found on both the \C\ islands and the bare Cu substrate after an additional sublimation of small amounts of \C\ onto the cooled sample.

A PtIr tip was attached to the free prong of a quartz tuning fork oscillating with a constant amplitude at its resonance frequency of $\sim 28\,\text{kHz}$.
The tip was further prepared {\em in-situ} by repeated indenting into the Cu substrate.
\C-tips were created by approaching the tip towards the molecule until it jumps to the tip~\cite{Schull2009}.
The bias voltage $V$ was applied to the sample.

Ab-initio calculations were performed with the DFT-FIREBALL code~\cite{Lewis2011} that uses the local density approximation (LDA) for the exchange and correlation potential, and a basis of numerical atomic orbitals (NAO) vanishing for a value higher than a cutoff radius.
For this kind of codes only the valence electrons are usually included in the calculation.
In this work, a spd basis for Cu and a sp for C are chosen with the corresponding cutoffs (given in Bohr radii): $r(Cu-s)=4.5$, $r(Cu-p)=5.7$, $r(Cu-d)=3.7$, $r(C-s)=4.5$, and $r(C-p)=4.5$~\cite{Schull2011}.
A Cu-slab of four layers with a $4\times4$ periodicity in the surface plane is built to simulate the metallic substrate (Figs.~\ref{Fig:Figure2}(a),(d)).
A single \C\ molecule is adsorbed on the slab and the system is relaxed using 16 wavevectors in the first Brillouin zone until the forces are lower than 0.05 eV/A.
The atoms of the lowest two metallic layers are kept fixed.
The tip is based on a four-layer Cu-pyramid oriented in the (111) direction (cf.~Ref.~\cite{svec_van_2012}), where the last Cu-apex is replaced by a \C\ molecule.
The atoms in the basis of the pyramid are kept fixed during the calculations.

The Cu pyramid tip with the \C\ attached is placed on the positions of a 17$\times$17 grid (spacing of 0.5\,\AA) above the adsorbed \C\ on the Cu slab.
Starting with the lowest atoms of the tip located 6\,\AA\ above the topmost atoms of the \C\ adsorbed on the surface, the tip was moved down in steps of 0.25\,\AA.
In every point all tip and sample atoms (except the fixed atoms mentioned above) were relaxed using the same conditions than that used for the isolated surface.
The vdW interaction was estimated by a semi-empirical correction based on the London expression~\cite{Ortmann2006}.
It is well known that the empirical values obtained from this expression for the C-C interaction are overestimated~\cite{Dappe2009}.
We use 60\% of the initial empirical value, which yields reasonable results as shown in Ref.~\cite{Gonzalez2009}.
The total energy with respect to the tip-surface distance is obtained by the sum of the DFT and vdW energies at each point of the grid.
Then, the forces are obtained by deriving the energy versus distance curves.
The force contribution without the vdW interaction does not exhibit an attractive regime (Fig.~S1).
This emphasizes the importance of vdW contributions as also recently shown for the adsorption of fullerenes~\cite{svec_van_2012}.
For this reason, the frequency shift $\Delta f$ was derived from the total tip--sample force $F_{TS}$ including vdW forces according to first-order perturbation theory~\cite{Giessibl1997}:
\begin{equation}
\label{eq:freq}
\Delta f (d)=\frac{f_{0}}{2\pi kA_{0}} \int_{0}^{2\pi} \! F_{TS}[d+A_{0}+A_{0}\cos\varphi]\,\cos\varphi \, \mathrm{d}\varphi.
\end{equation}

\noindent where the experimental values have been taken into account: $f_{0}$ is the resonance frequency (28\,kHz), $A_{0}$ the amplitude of oscillation (2\,\AA) and $k$ the stiffness of the AFM sensor ($\approx 1800$\,N/m).
The theoretical $\Delta f$ maps were generated from values calculated on the 17$\times$17 grid.
The used tip radius, which is important for vdW interaction, was varied from 1\,nm to 5\,nm, which influences the total value of $\Delta f$ while the general structures of the maps discussed below remain unchanged.

Calculated STM images were obtained using the same tip, following the methodology explained in Ref.~\cite{Blanco2004}.
Based on the non-equilibrium Green-Keldysh formalism, both systems tip and sample can be treated separately in the tunneling regime to be finally coupled by an interaction given by the hopping matrices $T_{ST}$ and $T_{TS}$, which denote the hopping of the electrons between sample and tip and {\em vice versa}.
The expression obtained for the electronic current in the tunneling regime at 0~K is written as:

\begin{equation}
\label{eq:tunnel}
I=\frac{4\pi e}{ \hslash} \int_{E_{F}}^{E_{F}+eV} \! Tr[T_{TS}\rho_{SS}(E)T_{ST}\rho_{TT}(E)] \, \mathrm{d}E.
\end{equation}

\noindent where $\rho_{SS(TT)}$ are the corresponding density of states (DOS) of the sample (tip) of the isolated system.

\section{Results and discussion}

First, small Cu clusters were deposited on the bare surface by approaching a metallic tip to the surface by a few Angstroms.
The lateral shapes and apparent heights of these Cu clusters imaged with a metallic tip (Fig.~S2) suggest that they consist of a few (presumably three) atoms.
Next, a \C\ molecule from the surface was transferred to the tip by approaching the tip sufficiently close (a few Angstroms from tunneling conditions).
The orientation of the molecule at the tip was determined by ''reverse imaging'' on small Cu clusters and showed that a hexagon of \C\ was facing the surface.

\begin{figure}[htb]
\centering
  \includegraphics[width=145mm]{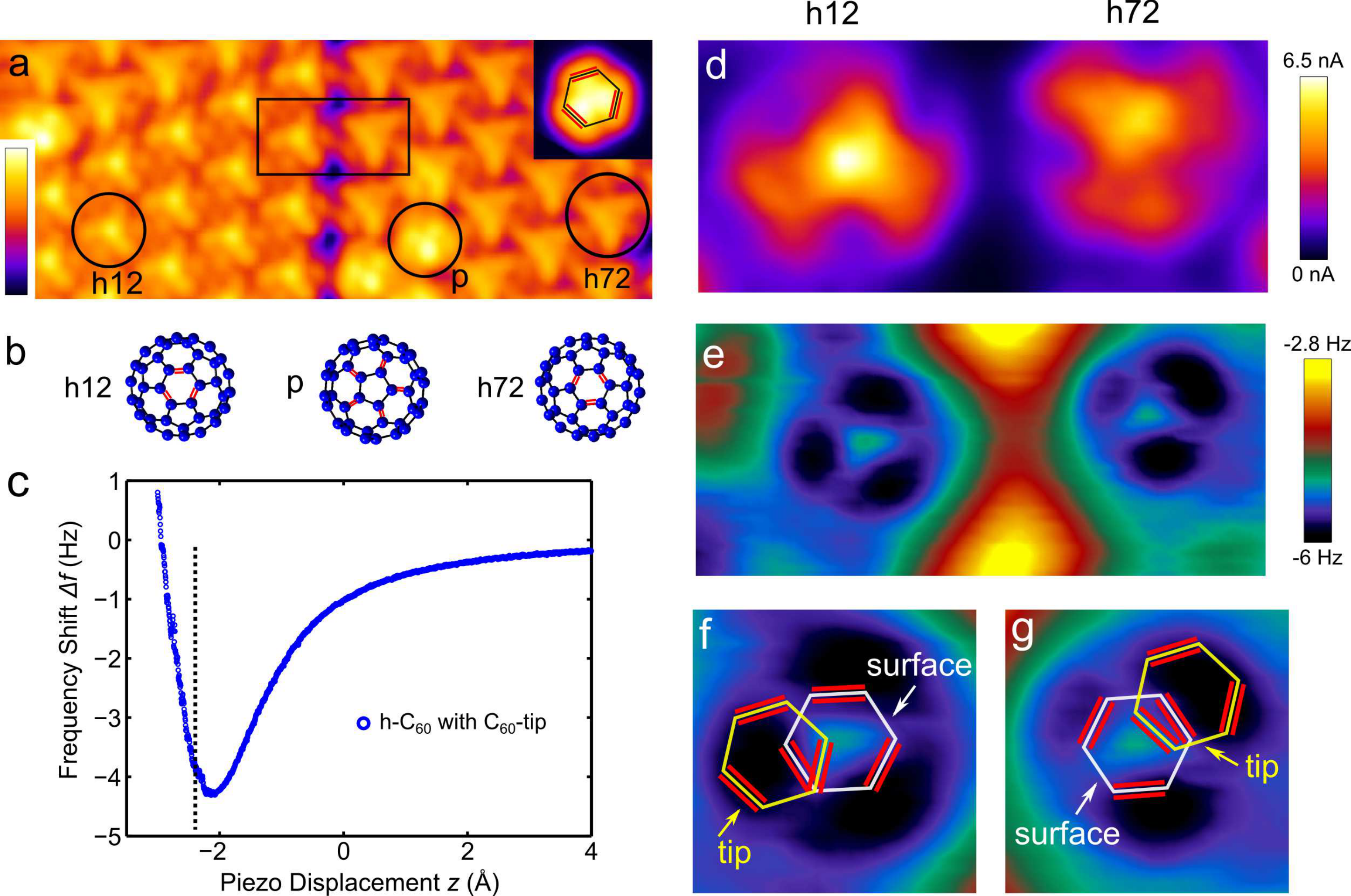}
 \caption{(a) Constant-current STM image ($V=1.7\,\text{V}$, $I=0.55\,\text{nA}$, image size: $10\times 4\,\text{nm}^2$) of a \C\ island obtained with a \C\ molecule attached to the tip apex (\C-tip).
  Three relative orientations denoted h12, p, and h72 are found (circles).
  Inset: Constant-current STM image ($V=-2\,\text{V}$, $I=0.36\,\text{nA}$, image size: $2\times 1.6\,\text{nm}^2$) of a copper cluster adsorbed on the surface imaged with the same \C-tip.
  The sketch shows the orientation of the hexagon of the \C\ at the tip that is closest to the surface.
	  (b) Sketches of the three orientations of \C\ on Cu(111) as viewed from the tip.
Red double-bars indicate the bonds between two hexagons.
  (c) Frequency shift $\Delta f$ versus piezo displacement $z$ measured above the center of a \hex.
  Simultaneously recorded maps of (d) the current $I$ and (e) frequency shift $\Delta f$ measured at $V=0.1\,\text{V}$ and at a constant tip-surface distance (dashed line) over h12 and h72 molecules with the \C-tip.
  The tuning fork oscillation was $A=(2.4\pm0.3)$\AA\@.
  (f) and (g): maps of $\Delta f$ of h12 and h72 molecules (same color scale than in (e)).
  The respective surface area is indicated by a black box in (a).
 The relative orientation of the uppermost hexagon of the surface \C\ and of the lowest hexagon of the molecule at the tip are overlaid.
 The sketch shows the arrangement of both hexagons for one of the positions of large attraction.}
  \label{Fig:Figure1}
\end{figure}
A constant-current image of a Cu cluster acquired with a \C-tip is shown in the inset of Fig.~\ref{Fig:Figure1}(a) ($V=-2\,\text{V}$).
At this voltage, electrons tunnel from the cluster into the second lowest unoccupied orbital (LUMO+1) of the molecule~\cite{Schull2009}.
The shape of the cluster reflects the spatial distribution of the LUMO+1 from which the orientation of the \C\ at the tip can be deduced.
Compared to typical STM images of the LUMO+1 using a metallic tip, the image of a \C\ molecule at the tip obtained with a metal cluster shows a mirror image of the molecule~\cite{Hauptmann2012}.
The sketch in the inset of Fig.~\ref{Fig:Figure1}(a) depicts the orientation of the surface-nearest hexagon of the \C\ molecule with the red bars indicating the bonds between two hexagons ([6,6] bonds), which are of higher bond order compared to bonds between a hexagon and a pentagon ([6,5] bonds)~\cite{Gross2012}.

\begin{figure}[htb]
\centering
  \includegraphics[width=145mm]{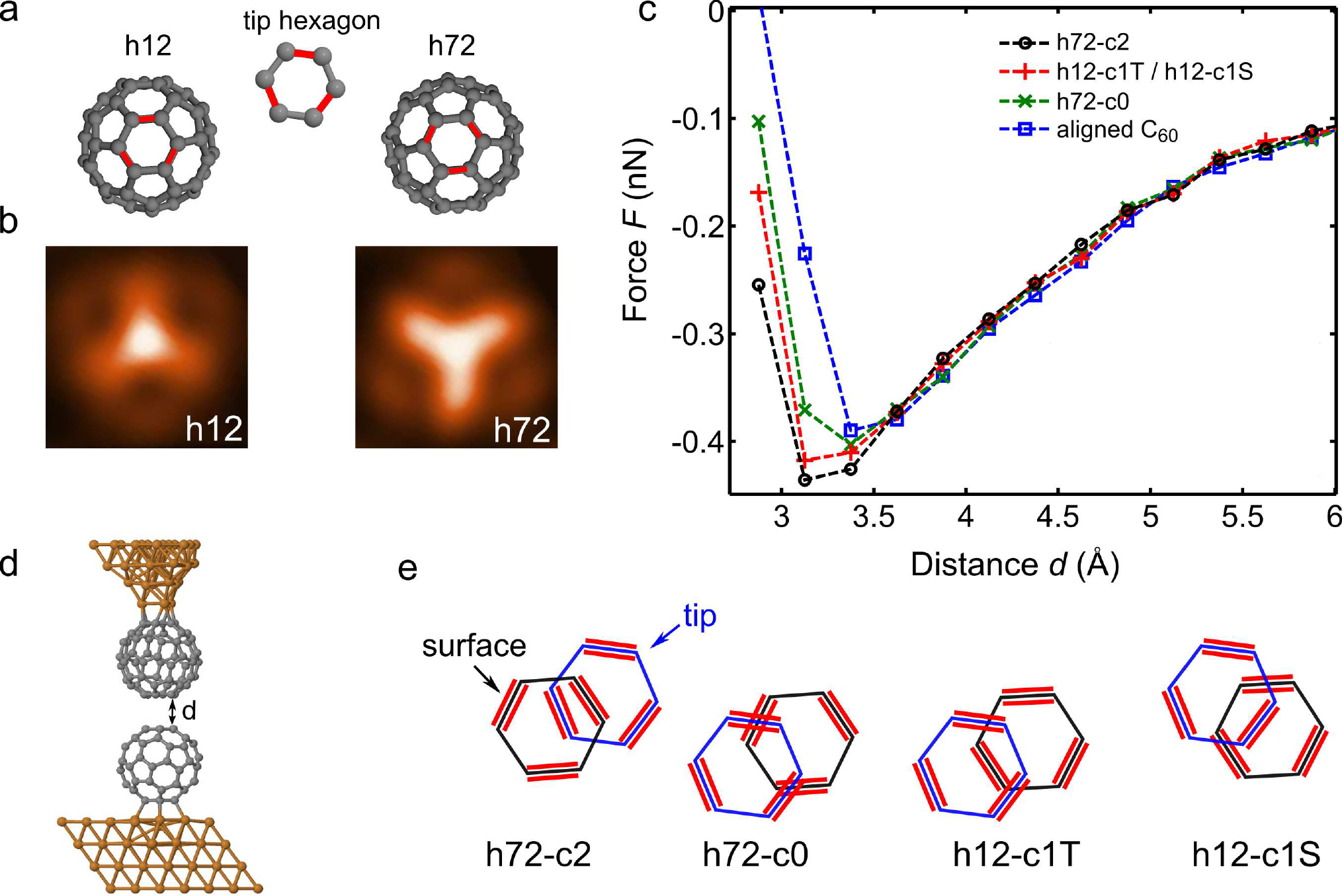}
 \caption{(a) Sketches of the upper halves of the \C\ molecules h12 and h72 as adsorbed on the Cu slab and the lowest hexagon of the \C--tip used for calculations. The red bars denote the [6,6] bonds.
 (b) Calculated STM images of h12 and h72.
 (c) Calculated force $F$ versus distance $d$ (defined in (d)) for different relative orientations of the \C\ at the tip and surface as defined in (e).
 (d) Sketch of the unrelaxed geometry for aligned \C\ molecules at tip and surface.
 (e) Sketches of the relative orientations of the uppermost hexagon of the \C\ at the surface and lowest hexagon of the \C-tip for the calculated force curves in (c).}
  \label{Fig:Figure2}
\end{figure}

A STM topograph of a \C\ island comprising two domains as imaged with the \C-tip is shown in Fig.~\ref{Fig:Figure1}(a).
The domains comprise molecules that either face with a hexagon (\hex) or a pentagon (\pen) towards the tip~\cite{Silien2004,Larsson2008}.
The different orientations of the molecules are shown in Fig.~\ref{Fig:Figure1}(b) with the [6,6] bonds sketched by red bars.
Different azimuthal rotations of the \hex\ molecules by 60\degree\ are present in the domains.
The top-most hexagon of the \hex\ (Fig.~\ref{Fig:Figure1}(a), inset) is rotated by either (12$\pm$6)\,\degree\ (referred to as h12) or (72$\pm$6)\,\degree\ (referred to as h72) with respect to the lowest hexagon of the \C-tip.
Depending on this relative orientation \hex\ appears either with a pronounced center (h12) or as a more homogenous structure (h72) in the STM image~\cite{Hauptmann2012}.
The threefold symmetry of the pentagon-hexagon ([5,6]) bonds of the \C-tip is clearly discernible on \pen .
As shown recently~\cite{Hauptmann2012,Lakin2013} as well as in Fig.~\ref{Fig:Figure2}(b), the symmetries of these patterns can be understood from a convolution of the local densities of electronic states (LDOS) of the tip and the sample.

Simultaneously acquired maps of the frequency shift $\Delta f$ and current $I$ of two \hex\ (black box in Fig.~\ref{Fig:Figure1}(a)) are shown in Figs.~\ref{Fig:Figure1}(d) and (e).
The feedback loop was switched off (at $V=1.7\,\text{V}$ and $I=0.55\,\text{nA}$) and a voltage of $V=0.1\,\text{V}$ was applied.
The tip-height was decreased until the frequency shift $\Delta f$ passes its minimum and increases again as indicated by the dashed line in the $\Delta f$ versus $z$ curve on a \hex\ (center position) in Fig.~\ref{Fig:Figure1}(c).

In both the I and $\Delta f$ map the molecules appear as almost three-fold symmetric patterns.
Given their broad spatial extension, the lobes in both maps approximately occur at the same spatial positions.
While the three lobes in the current map are subject to an increased tunneling current, the dark lobes in the $\Delta f$ map correspond to spatial positions with a minimal frequency shift.
These depressions indicate locations of strong attractive interaction.

To obtain a deeper understanding of the data, we analysed the relative orientation of the \C-tip and the molecules on the surface.
We concentrated on the positions of minimal frequency shift of the h12 and h72 as these lobes are more pronounced than the lobes in the current map (Fig.~\ref{Fig:Figure1}(d)).
For simplicity, we only discuss the orientations of the uppermost hexagon of the \C\ on the sample surface and the lowest hexagon of the \C\ at the tip~\cite{note2}.
In Figs.~\ref{Fig:Figure1}(f) and (g) the relative positions of these hexagons are shown for one of the three positions of minimal frequency shift for both the h12 and h72 orientations.
The red bars indicate the [6,6] bonds.
It turns out that the positions of minimal frequency shift occur when [6,6] bonds of the molecule at the tip and on the surface are facing a hexagon of the other \C.

Compared with recent measurements using a metallic tip~\cite{Pawlak2011} to image \C\ molecules in the constant height AFM mode, our observation of three clearly pronounced attractive lobes is fairly different.
Interestingly, a rather different contrast was found in maps of $\Delta f$ acquired with a CO-functionalized tip~\cite{Gross2012}:
The [6,6] bonds showed an increased $\Delta f$ and appeared shorter compared to the [6,5] bonds.
This contrast originates from strong Pauli repulsion of the [6,6] bonds, which have a higher bond order than the [5,6] bonds.
Here, we observe a different contrast mechanism giving rise to three well-pronounced attractive maxima, which are due to strong attraction of a [6,6] bond with a \C\ molecule.
The [6,6] bonds also play an important role for charge injection.
A recent work using a metallic tip to contact \C\ molecules showed that charge injection into \C\ molecules is more efficient at [6,6] bonds due to the formation of a bonding state of the tip apex atom and the \C\ molecule on the surface~\cite{Schull2011}.

Next, we corroborate our results by DFT+vdW calculations.
The used geometries for the h12 and h72 orientations~\cite{Note1} as well as the orientation of the lowest hexagon of the \C-tip are shown in Fig.~\ref{Fig:Figure2}(a).
Calculated STM images for both orientations (Fig.~\ref{Fig:Figure2}(b)) compare well with the experimental data (Fig.~\ref{Fig:Figure1}(a)).
Force ($F$, including vdW interaction) versus distance $d$ curves at different orientations and different lateral positions of the \C-tip above the \C\ on the surface are used to determine the distance range of interest (Fig.~\ref{Fig:Figure2}(c)).
As depicted in Fig.~\ref{Fig:Figure2}(d), $d$ is defined as the distance between the topmost carbon atoms from the \C\ on the surface and
the lowest atoms from the tip for the unrelaxed system.
The relative orientations of the uppermost hexagon of the \C\ at the surface and lowest hexagon of the \C-tip are defined in Fig.~\ref{Fig:Figure2}(e).
The force curves in Fig.~\ref{Fig:Figure2}(c) were calculated for different relative lateral positions of the two molecules: (i) a [6,6] bond of each molecule is facing a hexagon of the other molecule (h72-c2), (ii) a [6,6] bond of one molecule is facing a hexagon and a [6,5] bond of the other molecule is facing a hexagon (h12-c1T and h12-c1S), (iii) a [6,5] bond of each molecule is facing a hexagon (h72-c0), and (iv) the hexagons of both molecules are facing each other (aligned C60).
The minima in the force curves (maximal attraction) occur at $d = 3.4$~\AA, which is comparable to the minimum of calculated force curves between a CO tip and a \C\ molecule~\cite{Gross2012}.
Clear differences in the forces at different spatial positions of the \C-tip are observed for $d<3.5$\,\AA\@.
For molecules with facing hexagons (aligned \C, Fig.~\ref{Fig:Figure2}(d)) the force is less attractive than for the other orientations sketched in Fig.~\ref{Fig:Figure2}(e).
We note that the forces for the h12-c1T and h12-c1S orientations are virtually the same.
The calculated force is maximal for the h72-c2 orientation where the hexagons of both \C\ molecules are above a [6,6] bond of the other \C.
This is in agreement with the experimental findings.

\begin{figure}[htb]
\centering
  \includegraphics[width=80mm]{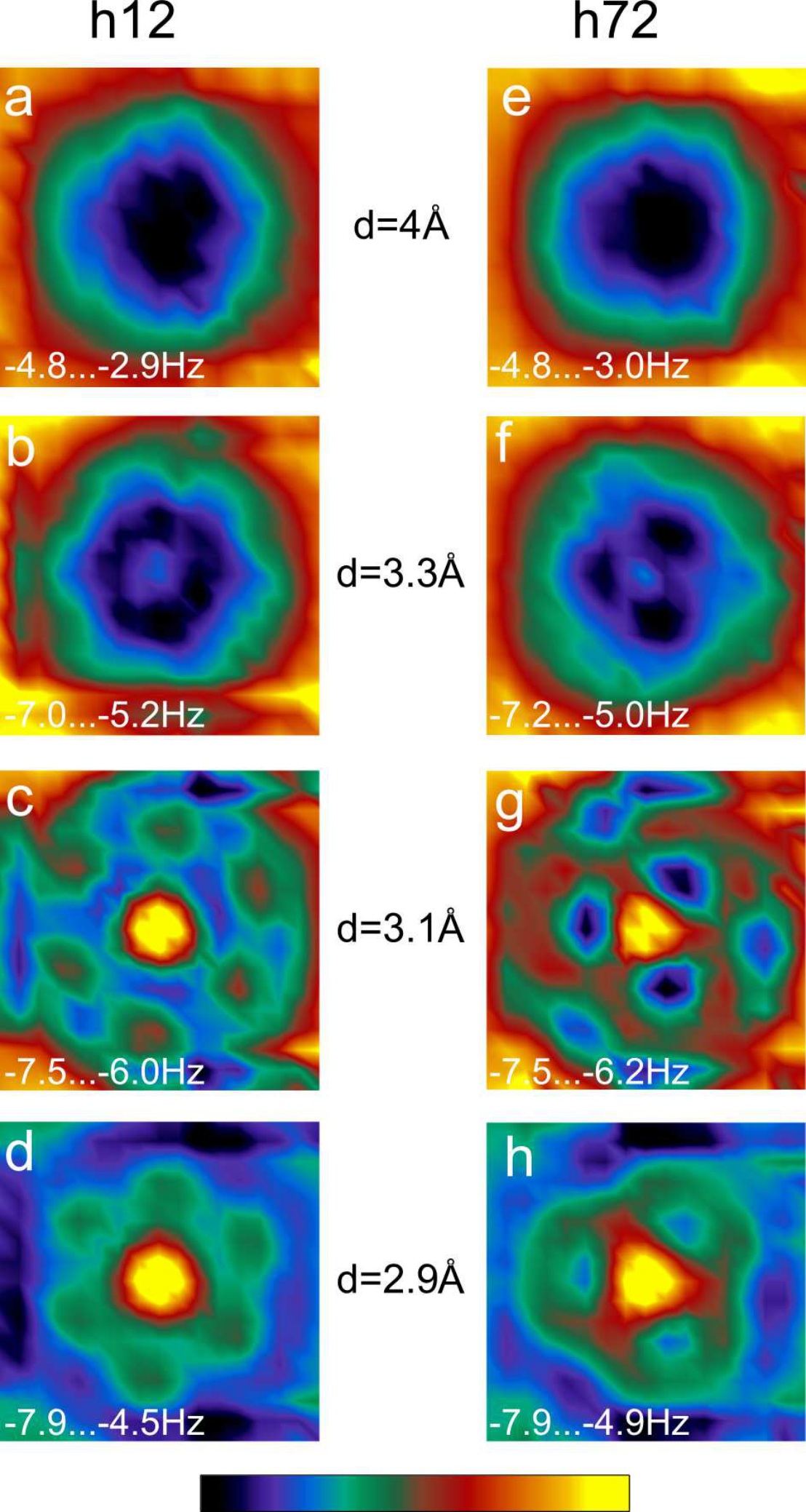}
  \caption{Calculations: Maps of $\Delta f$ for different $d$ of the h12 (a to d) and h72 (e to h) orientation.
  The color scale for all images is given at the bottom with the corresponding frequency shift margins indicated in every image.}
  \label{Fig:Figure3}
\end{figure}

Figures~\ref{Fig:Figure3}(a) to (h) show calculated maps of $\Delta f$ for the h12 and h72 orientations at four different distances.
For $d=4$\,\AA\ the maps are similar for both orientations h12 and h72 (Figs.~\ref{Fig:Figure3} (a) and (e)) showing a decrease of $\Delta f$ over the \C\ molecule.
At this distance regime the contrast is governed by long-range vdW interaction~\cite{Pawlak2011, Gross2009, Gross2012}.

For a further decrease of $d$, submolecular patterns arise.
In case of the h72 orientation, the $\Delta f$ map shows a threefold pattern of spatial maxima of attraction that occur at position where a [6,6] bond of one molecule faces a hexagon of the other molecule and vice versa.
In the center of the map, where surface and tip molecules are aligned, the calculation shows a reduced attractive interaction.
These results match very well with the experimental findings (Fig.~\ref{Fig:Figure1}(g)).

For the h12 orientation, the calculation results in a more six-fold symmetry at $d=3.3$\,\AA\ instead of the three lobes observed in the experimental data.
At $d=3.1$\,\AA\ a faint three-lobe structure can be discerned.
However, it is less pronounced than in the experimental data.
This may be due to parameters that our calculation does not account for, \eg, the lack of rotational relaxations or incomplete charge transfer from the \C\ to the metal atoms, and may be related with the almost negligible attraction between both \C\ molecules in the calculation (Fig.~S1).
A rotation of the \C\ molecule at the tip has been previously suggested to explain the different STM features observed when equally oriented \C\ molecules are imaged with a functionalized \C-tip~\cite{svec_van_2012}.
For $d<3$\,\AA\ the repulsive contribution increases as seen from Figs.~\ref{Fig:Figure3}(d) and (h).

\section{Summary}
Molecules can be used to define the structure of the tip.
CO-functionalized tips are often used and enable remarkable contrast in AFM and STM images.
Compared to those, \C\ tips are less flexible but the different types of molecular bonds strongly influence the conductance and interaction resulting in images with a complex intramolecular contrast.\\
We performed combined STM/AFM measurements on \C\ molecules using \C-functionalized tips.
The attractive interaction is maximal when a [6,6] bond of one molecule faces a hexagon of the other molecule and vice versa.
The experimental findings are corroborated by DFT+vdW calculations which show that the [6,6] bonds and hexagons exhibit a strong attractive interaction.

\section{Acknowledgments}
We thank Deutsche Forschungsgemeinschaft for financial support via the SFB 677 and for funding from the EU projects PAMS (agreement no. 610446) and the ERC Advanced Grant CEMAS (291194).

\providecommand{\newblock}{}


\begin{thebibliography}{10}
\bibitem{note2}
{The uppermost hexagon of the {C$_{\text{60}}$} molecule at the surface and the
  lowermost hexagon of the {C$_{\text{60}}$} molecule at the tip are the
  hexagons that are closest to each other according to Fig. 2(d).}
\end{thebibliography}
\end{document}